\documentclass[a4paper]{jpconf}
\usepackage{graphicx}
\usepackage{amssymb}
   \usepackage{amsmath}
    \usepackage{epsfig}
     \usepackage{bm}
\begin{document}
\title{Geometry of gauge links in operator definitions of transverse-momentum dependent PDFs: \\
A comparative view\footnote{Presented at the International Conference ``Photon-2011'', 22 - 27 May 2011 Spa, Belgium}}
\author{Igor O. Cherednikov$^{1,2}$}

\address{$^1$ Universiteit Antwerpen, Belgium}
\address{$^2$ BLTP JINR, Dubna, Russia}

\ead{igor.cherednikov@ua.ac.be}

\begin{abstract}
Different approaches to the problem of a gauge-invariant operator definition of transverse-momentum dependent parton densities (TMDs) are reviewed and compared from the point of view of their compatibility with the operator definition of collinear (integrated) parton distribution functions. In particular, geometry of the longitudinal and transverse gauge links (Wilson lines) both in unsubtracted quark TMDs and in the associate soft factors is addressed. A possible connection between different operator definitions is also discussed.
\end{abstract}

Classical deep inelastic $lH\,\,\to\,\,l' X$ scattering experiments deepened considerably our knowledge of nucleon structure. Measuring the momentum of the outgoing lepton $l'$ we learn about the distribution of partons inside the nucleon. In a reference frame where the nucleon moves (infinitely) fast, this information is accumulated in the parton distribution functions (PDFs) $f_a(x, Q^2)$  in terms of the partonic degrees of freedom: e.g., the Bjorken variable $x_{\rm Bj}$ relates to the fraction of the longitudinal momentum $P$ of the parent hadron possessed by a parton of the flavor $a$.
Such collinear (integrated) PDFs can be properly defined as completely gauge invariant (nonperturbative) hadronic matrix elements
(here and in what follows, the light-cone components of four-vectors are $p^\pm = (p^0 \pm p^z)/{\sqrt 2}$)
\begin{equation}
  f_a(x, \mu^2)
=
  \frac{1}{2}
  \int \frac{d\xi^- }{2\pi }
  \ {\rm e}^{-ik^{+}\xi^{-} }
  \left\langle
              p\  |\bar \psi_a (\xi^-, \mbox{\boldmath$0_\perp$})[\xi^-, 0^-]_n
              \gamma^+
   \psi_a (0^-,\mbox{\boldmath$0_\perp$}) | \ p
   \right\rangle \
   \label{eq:iPDF}
\end{equation}
with renormalization-group properties controlled by the DGLAP evolution equations (for review and Refs. see \cite{DIS_p}). The Wilson line $[\xi^-, 0^-]_n$ will be introduced below.
Moreover, one can relate the moments of the collinear PDFs
$
M^N_a = \int\!dx \ x^{N-1} f_a(x)
$
to the matrix elements of the local twist-two operators
$
{\cal O}^N = (p^+)^{-N} \ \langle  p \ | \frac{1}{2} \bar \psi_a (0) \{ \gamma^+ iD^+...iD^+ \}_{\rm sym.} \psi_a(0) | \ p \rangle
$
arising in the operator product expansion on the light-cone, thus making them well-defined objects from the field-theoretical point of view \cite{CS_first}.


Another attractive feature of the collinear PDFs (\ref{eq:iPDF}) is that they imply a clear interpretation as the probability for a parton inside the nucleon to have the longitudinal momentum $k_{\rm long.} = xP_{\rm long.}$.
This interpretation is most naturally established when QCD is canonically quantized (and subsequently renormalized) on equal-``light-cone-time'' surfaces $\xi^+ = 0$ in a class of singular non-covariant gauges \cite{LC_quant}.
This parton number interpretation holds, in the light-cone gauge $A^+ =0$, in higher order calculations as well \cite{LC_Bassetto, SF87, CDL84}.
However, an important problem must be solved in order to make the calculations in the light-cone gauge feasible.
Even after imposing the gauge condition $A^+=0$, the gauge in not completely fixed: one may still perform a $\xi^-$-independent gauge transformation $U(\xi^+, \bm{\xi}_\perp)$ by virtue of
$
\partial^+ U(\xi^+, \bm{\xi_\perp})
=
\partial / \partial x^- U(\xi^+, {\bm \xi_\perp})
= 0
$.
Therefore, fixing the gauge is equivalent to imposing certain boundary conditions on the gauge field (see, e.g., \cite{SF87, TMD_LC_trans, BR05, AT_POL}).  Going over to the conjugate momentum space, one observes that the ambiguity in the behavior of the gauge field at light-cone infinity $\xi^- \to \infty$ maps over ambiguity of the gluon Green function at small $q^+ \to 0$.
A key issue is, therefore, how to get rid of extra complications due to the emergent ``spurious'' singularity $\sim [q^+]^{-1}$ in the free gluon propagator
\begin{equation}
 D^{\mu\nu} (q)
 =
 \frac{i}{q^2+i0} \left( - g^{\mu\nu} + \frac{(n^-)^{\mu} q^\nu + (n^-)^{\nu} q^\mu}{[q^+]} \right) \ .
\label{eq:gluon_pr_LC}
\end{equation}
The uncertainty of the pole prescription in Eq. (\ref{eq:gluon_pr_LC}) corresponds to the residual gauge freedom and can be treated without changing the gauge-fixing constraint $A^+=0$.

There are several possible pole-prescription-fixing procedures that are compatible with the light-cone gauge and have been shown to give correct results (at least, up to the $O(\alpha_s^2)$-order). In particular, the principal value prescription
$
\frac{1}{[q^+]_{\eta}^{\rm PV} }
       =  \lim_{\eta \to 0}\ \frac{1}{2}\left( \frac{1}{q^+ + i \eta}
                             +\frac{1}{q^+ - i \eta} \right)
$
was used in Ref. \cite{CFP80} to evaluate the DGLAP kernel in the next-to-leading order. Non-symmetrical advanced and retarded pole prescriptions are also possible \cite{TMD_LC_trans, BR05}.
Although these methods work in some situations, the only pole prescription which is consistent with the equal-time canonical quantization in the light-cone gauge is the Mandelstam-Leibbrandt one \cite{LC_ML}:
\begin{equation}
   \frac{1}{[q^+]} \to  \lim_{\eta \to 0}\ \frac{1}{[q^+]_{\rm ML }}
       =
       \lim_{\eta \to 0}\  \frac{(q \cdot n^+)}{(q\cdot n^+) (q\cdot n^-) + i \eta}
       \doteq
       \lim_{\eta \to 0}\  \frac{1}{ (q\cdot n^-) + i \eta (q \cdot n^+) }\ ,
       \label{eq:ML_def}
\end{equation}
where and in what follows $n^\pm$ are the light-like vectors $(n^\pm)^2 = 0 \ , \ n^+ n^- = 1$, and $\doteq$ means equality in the sense of the theory of distributions.
It was shown that the free gluon propagator supplied with the ML pole prescription can be directly derived following the equal-time quantization procedure and is compatible with well-established results at least up to $O(\alpha_s)$-order \cite{LC_Bassetto, BR05, K_LC}. The main difference between the $q^-$-independent prescriptions and the ML one (\ref{eq:ML_def}) originate in the different situation of poles in the $q^0$ plane, as it is shown and explained in Fig. 1. Thanks to this feature, one can perform a Wick rotation of the integration contour to the Euclidean momentum space,  and the ultraviolet divergences can be analyzed by means of the usual power counting procedure in the Euclidean space. This observation  anticipates the absence of overlapping divergences in the loop calculations with the ML prescription.
\begin{figure}[h]
\begin{center}
\includegraphics[width=0.30\textwidth,height=0.40\textheight,angle=90]{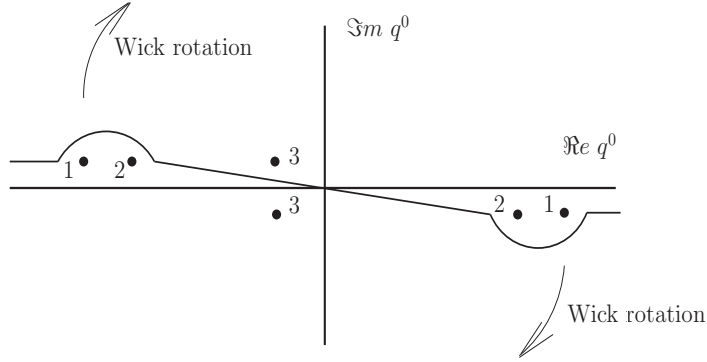}\hspace{2pc}%
\caption{\label{fig:2} Location of the poles in the complex $q^0$ plane: the poles of the light-cone gluon propagator with the ML prescription (1) and the poles of the propagator in a covariant gauge (2) belong to the same quadrants, so that the clock-wise Wick rotation is allowed. The poles of the light-cone propagator with the principal-value prescription (3), in contrast, impede that rotation.}
\end{center}
\end{figure}

In is worth noting that albeit the pole-prescription issues mentioned above may reveal themselves in the course of the calculation, they are non-visible in the case of the collinear PDFs, by virtue of the cancelation of the soft divergences in the virtual and the real gluon exchange graphs. However, those issues are crucial and unavoidable in unintegrated PDFs \cite{CS_first}, which are introduced in the factorization approach to the semi-inclusive processes. The picture of the nucleon revealed in the DIS experiments, being essentially one-dimensional, is still incomplete: in this ``longitudinal'' picture the transverse degrees of freedom of the partons are eliminated by definition and the 3D-structure remains inaccessible. The study of semi-inclusive processes, such as semi-inclusive deep inelastic scattering (SIDIS), the Drell-Yan (DY) process,  hadron-hadron collisions, or lepton-lepton annihilation to hadrons, where (at least) one more final or initial hadron is detected and its transverse momentum (and, possibly, its spin) is observed, calls for the introduction of more involved quantities---unintegrated transverse-momentum dependent (TMD) distribution and fragmentation functions (see \cite{TMD_INT} and Refs. therein). Moreover, a variety of applications of the TMDs approach has been found in the phenomenology of polarized hadronic processes. It was conjectured and corroborated that the idea of TMD parton densities inside the nucleons is directly applicable in the theory of single-spin asymmetries (see, e.g.,  \cite{TMD_PHENO} and Refs. therein). Recently, interesting possibility that linearly polarized gluons inside unpolarized protons can affect the cross-sections of the scalar and pseudoscalar Higgs boson production in the gluon-gluon fusion channel was proposed and discussed within the TMD approach in the Refs. \cite{TMD_LHC}.

In the present paper, I give a brief account of how the problem of the emergent singularities beyond the tree-approximation  is approached in different operator definitions of the (quark) TMDs. At the one-loop level, the following three classes of singularities  are expected: $(i)$ simple ultraviolet poles which must be removed by the standard renormalization procedure; $(ii)$
pure rapidity divergences, which depend on an additional rapidity parameter, but do not jeopardize the renormalizability of the TMDs, and can be safely resumed by means of the Collins-Soper equation; $(iii)$ highly undesirable overlapping divergences: they contain the UV and rapidity poles simultaneously and thus break down the standard renormalizability of TMDs, calling for a {generalized} renormalization procedure in order to enable the construction of a consistent operator definition of the TMDs. Before getting started with the analysis of the divergences, let us try to learn something from the tree-approximation.
The simplest ``unsubtracted'' definition of ``a quark in a quark'' TMD ($[{\rm A}]_{\rm n}$ means that we use the light-like longitudinal Wilson lines), which allows a {\it parton number interpretation} in the light-cone gauge, reads
\begin{eqnarray}
&& {\cal F}_{\rm unsub.}^{[{\rm A}_{\rm n}]} \left(x, {\bm k}_\perp; \mu \right)
=
  \frac{1}{2}
  \int \frac{d\xi^- d^2 {\xi}_\perp}{2\pi (2\pi)^2} \
  {\rm e}^{-ik \cdot \xi}
  \left\langle
              p \ |\bar \psi_a (\xi^-,  \bm{\xi}_\perp)
              [\xi^-,  \bm{\xi}_\perp;
   \infty^-,  \bm{\xi}_\perp]_{n}^\dagger  \right.  \nonumber \\
   && \left.
\times
   [\infty^-,  {\xi}_\perp;
   \infty^-,  {\infty}_\perp]_{\bm l}^\dagger
   \gamma^+[\infty^-,  {\infty}_\perp;
   \infty^-, \bm{0}_\perp]_{\bm l}
   [\infty^-, \bm{0}_\perp; 0^-,\bm{0}_\perp]_{n}
   \psi_a (0^-,\bm{0}_\perp) | \ p
   \right\rangle \
\label{eq:general}
\end{eqnarray}
with ${\xi^+=0}$.
Generic semi-infinite path-ordered gauge links evaluated along a given four-vector $w$ are defined as
$
[\infty; \xi]_{w}
\equiv {}
  {\cal P} \exp \left[
                      - i g \int_0^\infty d\tau \ w_{\mu} \
                      A_{a}^{\mu}t^{a} (\xi + w \tau)
                \right] \ ,
$
where, in the case under consideration, the vector $w$ can be either light-like $w_L = n^\pm\ , \ (n^\pm)^2 =0$, or transverse $w_T = {\bm l}$. Although the formal integration of
definition (\ref{eq:general}) over $\bm k_\perp$ yields the collinear PDF, Eq. (\ref{eq:iPDF}),
\begin{equation}
  \int\! d^2 \bm k_\perp \ {\cal F}_{\rm unsub.}^{[{\rm A}_{\rm n}]} (x, \bm k_\perp)
  =
 \frac{1}{2}
  \int \frac{d\xi^- }{2\pi } \
  {\rm e}^{-ik^{+}\xi^{-} } \
  \left\langle
              p\  |\bar \psi_a (\xi^-, \bm 0_\perp)[\xi^-, 0^-]_n
              \gamma^+
   \psi_a (0^-,\bm 0_\perp) | \ p
   \right\rangle \ = f_a(x) \ ,
   \label{eq:u_to_i}
\end{equation}
this is only well-justified in the tree approximation, because the rapidity divergences in the loop corrections prevent a  straightforward reduction.
I would like to emphasize that the normalization of the above TMD
\begin{equation}
{\cal F}_{\rm unsub.}^{[{\rm A}_{\rm n}] (0)} (x, {\bm k}_\perp)
=
  \frac{1}{2}
  \int \frac{d\xi^- d^2
 \bm{\xi}_\perp}{2\pi (2\pi)^2}
  {\rm e}^{- i k^+ \xi^- + i  \bm{k}_\perp \cdot  \bm{\xi}_\perp}
 { \langle p \ | }\bar \psi (\xi^-,  \bm{\xi}_\perp)
   \gamma^+
   \psi (0^-, \bm 0_\perp) { | \ p \rangle } =
   \delta(1 - x ) \delta^{(2)} (\bm k_\perp) \
\label{eq:tree_tmd}
\end{equation}
can be obtained following the {canonical quantization procedure in the light-cone gauge}, where the longitudinal Wilson lines disappear and the equal-time commutation relations for the quark creation and annihilation operators $\{a^\dag (k, \lambda), a(k, \lambda)\}$ lead immediately to the parton number interpretation of the TMD:
\begin{equation}
 {\cal F}_{\rm unsub.}^{[{\rm A}_{\rm n}] (0)} (x, {\bm k}_\perp)
 \sim
 \langle\  p \  | \ a^\dag(k^+, \bm k_\perp; \lambda) a(k^+, \bm k_\perp; \lambda) \ | \ p\ \rangle
 \ .
 \label{eq:parton_N}
\end{equation}
Use of the ``tilted'' gauge links doesn't obey this requirement.


\begin{figure}[h]
\includegraphics[width=0.45\textwidth,height=0.70\textheight,angle=90]{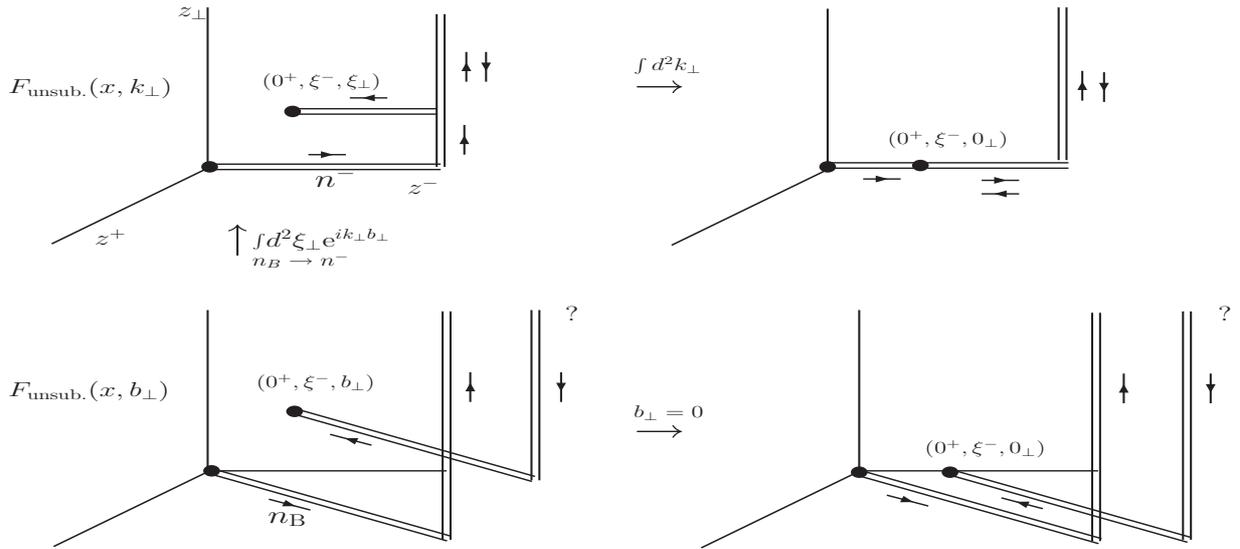}\hspace{2pc}%
\caption{\label{fig:1} Geometrical structure of integration contours in the unsubtracted TMDs with the light-like (upper panel) and off-the-light-cone (lower panel) longitudinal gauge links and their symbolic reduction to the collinear PDFs. In the former case, the transverse gauge links cancel completely after the $\bm k_\perp$-integration, while the longitudinal gauge links reduce to the one-dimensional light-like connector $[\xi^-, 0^-]$. In the off-the-light-cone situation, the cancelation of the transverse gauge links at infinity is not, at least, straightforward. Moreover, the integrated configuration contains two non-vanishing off-the-light-cone gauge links which are not equivalent to the simple connector $[\xi^-, 0^-]$. Beyond the tree-level, the renormalization group properties of those two objects are also different. I put the interrogation marks next to the transverse gauge links at infinity since I'm not aware of any consistent treatment of them in the TMD formulations with off-the-light-cone (tilted) Wilson lines. In contrast, the transverse gauge links appear naturally in the ``light-cone'' frameworks.}
\end{figure}

\begin{figure}[h]
\includegraphics[width=0.45\textwidth,height=0.70\textheight,angle=90]{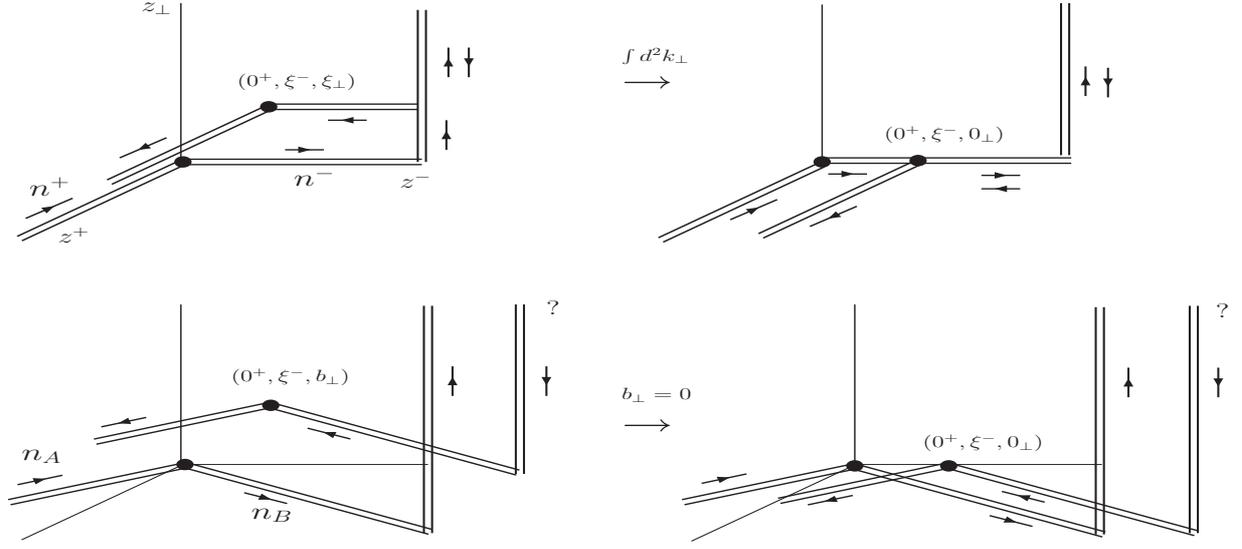}
\caption{\label{fig:2}Comparative geometry of the Wilson lines in unsubtracted soft factors and visualization of the reduction to the collinear case. Upper panel shows the soft factor in the momentum space, as proposed in the Refs. \cite{CS_all, CS09}. Lower panel presents the tilted off-the-light-cone integration paths in the impact parameter space, as well as the result of the reduction to the collinear $\bm b_\perp \to 0$ configuration.}
\end{figure}

\begin{figure}[h]
\includegraphics[width=0.45\textwidth,height=0.70\textheight,angle=90]{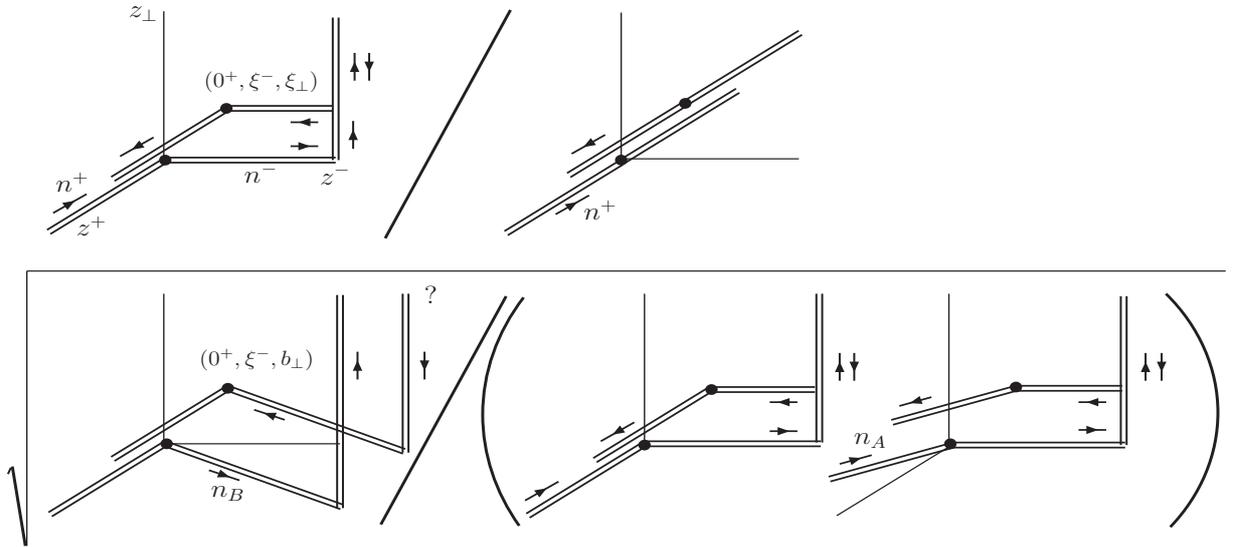}
\caption{\label{fig:2}Comparative geometry of the Wilson lines in the subtracted soft factors. Upper panel corresponds to the soft factor of the TMD distribution function which enters the factorization formula (\ref{eq:LC_factor}). Lower panel shows the longitudinal gauge links shifted off the light-cone, which are used in the factorization approach (\ref{eq:Col_factor}).}
\end{figure}

Going beyond the tree-approximation, one encounters a bunch of singularities mentioned above.
To overcome the problems related with them, different frameworks have been proposed.

Adopting the covariant Feynman gauge, Ji, Ma and Yuan developed a framework which makes use of the tilted (off-the-light-cone) longitudinal gauge links lined up along the vector $n_B^2 \neq 0$ \cite{JMY}. In a covariant gauge, the transverse gauge links at light-cone infinity cancel, and the rapidity cutoff $\zeta = (2 p\cdot n_B)^2/|n_B^2|$ is introduced to control the deviation of the longitudinal gauge links from the light-like direction. A subtracted soft factor contains the non-light-like gauge links as well. Within this approach, the off-the-light-cone unsubtracted TMDs where the light-like vector $n^-$ in the longitudinal Wilson lines is replaced by the tilted vector $n_{\rm B} = (-{\rm e}^{2y_B}, 1, \bm 0_\perp)$ do not satisfy the relation (\ref{eq:u_to_i}), even in the tree approximation---cf. Fig.~2 and the caption. However, one can design a ``secondary factorization'' method which allows the expression of this TMD (transformed to the impact parameter space ${\cal F} (x, \bm b_\perp)$) in terms of a convolution of collinear PDFs and perturbative coefficient functions at small $\bm b_\perp$ \cite{JMY}.
Another subtraction method, also in covariant gauges, but without explicit off-the-light-cone regularization in the unsubtracted TMD, was developed in Refs. \cite{Collins_what, TMD_subtract}. The corresponding geometry of the light-like and tilted Wilson lines in the soft factors is shown in Fig. 3, lower panel.

In our works \cite{CS_all} we proposed to study the renormalization-group properties of the unsubtracted quark TMD (\ref{eq:general}) and to make use of its one-loop anomalous dimensions in order to reveal the simplest {\it minimal} geometry of the gauge links in the soft factor which allows one to get rid of the mixed rapidity-dependent terms. We showed (in the leading $O(\alpha_s)$-order) that the extra contribution to the anomalous dimension is nothing but the cusp anomalous dimension \cite{KR87, K_LC}. Note that in these works we adopted the light-cone gauge supplied with the $q^-$-independent pole prescriptions. In subsequent works we showed that our approach works in the case of the ML pole prescription as well, Refs. \cite{CS09}, and that it can be consistently used to formulate a generalized definition of the quark TMD with a {\it non-minimal spin-dependent term} in the Wilson lines, Refs. \cite{CKS10} (see also \cite{CDKM06}).

Compared to the Ji-Ma-Yuan approach, we followed a different strategy. Making the assumption that the parton number interpretation (\ref{eq:parton_N}) must hold for TMDs in the light-cone gauge (like it holds in the collinear PDFs), we are in a position to {\it derive} a gauge-invariant operator definition of the TMD. In other words, starting from the requirement of the probability interpretation in the light-cone gauge and adding, step by step, the {\it minimally necessary} gauge links, we would end up with a gauge invariant operator definition of the TMD without undesirable singularities \cite{Ste83}. The generalized definition of the quark TMD we proposed, reads
\begin{equation}
  {\cal F}^{[{\rm A}_{\rm n}]} (x, \bm k_\perp; \mu, \theta)
  =
\frac{{\cal F}_{\rm unsub.}^{[{\rm A}_{\rm n}]} (x, \bm k_\perp; \mu, \theta)}{S_F(n^+,n^-; \theta) L_F^{-1}(n^+)} \ ,
\label{eq:TMD_LC}
\end{equation}
where the soft factor $S_F$ and the self-energy factor $L_F$ are defined in Refs. \cite{CS_all, ChSt_edge}, see also Figs.~3 and~4. The rapidity regulator is $\theta = \frac{n^+n^-}{\eta}$.

Let me suggest the following {\it conjecture} concerning the generic structure of divergences in (any reasonable) operator definition of the TMD beyond the tree-approximation:

The contribution of the overlapping $\sim 1/\epsilon \otimes \ln \theta$ singularities to the renormalized TMD can be expressed either in terms of a finite number of the cusp anomalous dimensions which are known in the theory of Wilson lines/loops up to the $O(\alpha_s^2)$-order---in this case, their treatment consists of the subtraction of (a finite number of) corresponding {\it cusped soft factors}; or those singularities depend on the {\it degenerate rapidities} $\sim \ln \theta_0 = \ln (n^\pm)^2$---in that case, one has to subtract the self-energy soft factors which consist, in contrast, of the ``smooth'' infinite gauge links without any obstructions (cusp or intersections). The conjecture is, therefore, that {\it there is no other sort of unphysical singularities in the loop corrections to the TMDs in any order of $\alpha_s$}.

In the leading order, we demonstrated the validity of the above statement in our works \cite{CS_all, CS09, CheISMD, ChSt_edge}. Recently, some interesting aspects of the TMD factorization, light-like Wilson lines and renormalization properties of the PDFs have been studied adopting the methodology of soft-collinear effective theory (SCET) in Refs. \cite{SCET_TMD}.
An important question remains, however, how our TMD can be built into an appropriate factorization formula for semi-inclusive hadronic tensor. In our approach, the following factorization formula is supposed to be valid
\begin{equation}
    W^{\mu\nu}
 =
 |H(Q,\mu)^2|^{\mu\nu} \cdot \frac{{\cal F}_{\rm unsub.}^{[{\rm A}_{\rm n}]} (x, \bm k_\perp; \mu, \theta)}{S_F(n^+,n^-; \theta) L_F^{-1}(n^+)} \otimes \frac{{\cal D}_{\rm unsub.}^{[{\rm A}_{\rm n}]} (z, z \bm k'_\perp; \mu, \theta)} {S_D(n^+,n^-; \theta) L_D^{-1}(n^-)} + ... \ ,
 \label{eq:LC_factor}
\end{equation}
where the geometrical structure of the soft factor is consistent with that of the collinear PDF \cite{K_LC, Li_97} and doesn't break the number interpretation. The explicit proof of the conjectured factorization and of absence of double-counting is in progress.

Recently, Collins proposed a new definition of the (quark) TMD \cite{New_TMD_Col} which is built into the factorization formula for the semi-inclusive hadronic tensor (up to power corrections)
\begin{equation}
 W^{\mu\nu}
 =
 |H(Q,\mu)^2|^{\mu\nu} \cdot {\cal F}^{\rm [Col.]} (x, \bm k_\perp; \mu, \zeta_F)\otimes {\cal D}^{\rm [Col.]} (z, z \bm k'_\perp; \mu, \zeta_D) + ... \ ,
 \label{eq:Col_factor}
\end{equation}
where all soft factors are absorbed into the TMD distribution ${\cal F}^{\rm Col.}$ and the fragmentation ${\cal D}^{\rm Col.}$ functions, so that there are no separate soft factors in factorized structure functions, e.g.,
\begin{equation}
  {\cal F}^{\rm [Col.]} (x, \bm b_\perp; \mu, \zeta_F)
  =
  {\cal F}_{\rm unsub.}^{[{\rm A}_{\rm n}]} (x, \bm b_\perp; \mu) \cdot \sqrt{\frac{S(n^+,n_B)}{S(n^+,n^-)S(n_A,n^-)}} \ .
\label{eq:TMD_Col}
\end{equation}
Here the soft factors depend on the light-like $n^\pm$ or the tilted $n_{A,B}$ vectors (for details, see \cite{New_TMD_Col}). Note that the TMD (\ref{eq:TMD_Col}) is defined in the impact parameter space ${\cal F} (x, \bm b_\perp) = \int\! d^{2} \bm k_\perp \ {\rm e}^{- i \bm k_\perp \bm b_\perp} \ {\cal F} (x, \bm k_\perp)$, so that it is, in fact, a ``semi-integrated PDF'' and the reduction to the collinear case corresponds to the limit $\bm b_\perp \to 0$. Some phenomenological and lattice applications of this approach have already been discussed in Ref. \cite{New_TMD_PHENO}. The geometry of the gauge links in the soft factors is presented and explained in Fig. 4.

Several open questions are still to be answered:
$(i)$ How should one prove the complete gauge invariance of the TMD (\ref{eq:TMD_Col})? It is formulated in the covariant Feynman gauge where the transverse gauge links at light-cone infinity vanish. What will change, if we adopt some physical (axial) gauge?
$(ii)$ In particular, how one should treat the $T-$odd effects in the axial gauges given that the structure of the transverse gauge links at light-cone infinity is not yet clarified in the TMD (\ref{eq:TMD_Col})?
$(iii)$ After reduction to the collinear PDF (in the case of TMD (\ref{eq:TMD_Col}), this corresponds to the limit $\bm b_\perp \to 0$), there is neither a mutual compensation of the longitudinal, nor that of the transverse gauge links (if introduced in the usual manner), see Fig. 4. Hence, the geometrical structure of the gauge links in the collinear PDF obtained from the TMD (\ref{eq:TMD_Col}) seems too cumbersome to be simply included in the standard DIS factorization scheme (see, e.g., Ref. \cite{K_LC, Li_97}).

To conclude, several approaches to the problem of the factorization of semi-inclusive processes which make use of the unintegrated TMD parton distribution functions have been proposed and developed so far. There is no {\it a priori} clear relationship between these frameworks: e.g., corresponding operator definitions of the TMDs may, in principle, describe different objects with different (renormalization-group, gauge invariance, evolution) properties. It is a matter of further study to work out these issues.

\paragraph{Acknowledgements}
The results presented in this work have been obtained in collaboration with N. G. Stefanis. I'm also grateful to him for careful reading of the manuscript and fruitful discussion of its content. I thank the Organizers of the conference Photon-2011 in Spa for the hospitality and warn atmosphere during the conference. The figures were produced with \verb"JaxoDraw" \cite{JaxoDraw}

\end{document}